\documentclass[aps,prl,twocolumn,letter]{revtex4-1}
\usepackage[english]{babel}
\usepackage{subeqnarray}
\usepackage{hyperref}
\usepackage{graphicx}
\usepackage{mathrsfs}
\usepackage{bm}
\usepackage{amssymb}
\usepackage{amsmath}
\usepackage{bbold}

\newcommand{\vv}[1]{\mathbf{#1}}
\newcommand{\bra}[1]{\left<#1\right|}
\newcommand{\ket}[1]{|#1\rangle}

\newcommand{\eq}[1]{(\ref{eq:#1})}
\newcommand{\fig}[1]{Fig. \ref{fig:#1}}
\newcommand{\abs}[1]{\left|#1\right|}
\begin{document}
\title{Dephasing of multiparticle Rydberg excitations for fast entanglement generation}
\date{\today}
\author{F. Bariani}
\author{Y.O. Dudin}
\author{T.A.B. Kennedy}
\author{A. Kuzmich}
\affiliation{School of Physics, Georgia Institute of Technology, Atlanta, GA, 30332-0430, USA}
\begin{abstract}
An approach to fast entanglement generation based on Rydberg dephasing of collective excitations
(spin waves) in large, optically thick atomic ensembles is proposed. Long range $1/r^3$ atomic interactions
are induced by microwave mixing of opposite-parity Rydberg states. Required long coherence times are achieved
via four-photon excitation and read-out of long wavelength spin waves. The dephasing mechanism is shown to have
favorable, approximately exponential, scaling for entanglement generation.
\end{abstract}
\maketitle
Alkali atoms excited to Rydberg levels are attracting increasing attention as candidates for quantum computation on the MHz scale \cite{saffman_review}.
Various protocols for quantum computation and multiparticle entanglement using Rydberg level interactions have been proposed in recent years \cite{saffman_molmer_2009, moller_2008,*muller_2009,*zhao2010}. These proposals rely on variations of the Rydberg blockade mechanism, where the presence of an excited Rydberg atom prevents (blocks) another atom, or atoms, from being excited \cite{duan2000,lukin2001}. This approach has already been used to generate entanglement of pairs of Rb atoms \cite{saffmannature,grangiernature}. The Rydberg blockade mechanism is in principle also applicable to create entanglement of collective excitations (spin waves), provided sufficiently small atomic ensembles are employed \cite{lukin2001,saffman2002}.
This attractive capability could permit realizations of large scale, complex entangled matter-light systems. The basic requirement of large optical thickness of the atomic ensembles is, however, in conflict with the short range of the blockade radius. The challenge remains to achieve sufficient optical depth with small ensembles ($<10$ $\mu$m) using tight, densely populated optical lattices and/or optical cavities.

In this Letter we propose an alternative approach that alleviates the difficulties of the small sample-blockade mechanism and makes it possible to realize fast entanglement generation and distribution in large, free-space atomic ensembles.
Rather than trying to prevent multiple excitations via the Rydberg blockade mechanism, our idea is to allow multiple Rydberg level excitations to self-interact and dephase.
The interaction-induced phase shifts suppress the contribution of multiply excited states in phase matched optical retrieval. The dephasing mechanism therefore permits isolation and manipulation of individual spin wave excitations.

The strong interaction required to dephase multiple excitations is induced by mixing adjacent, opposite-parity Rydberg levels with a microwave field \cite{adamsmuwave}. These levels experience resonant dipole-dipole interactions ($ns + n'p \rightarrow n'p + ns$) that extend over the whole ensemble in contrast to the weaker, short range Van der Waals coupling due to non-resonant processes ($ns + ns \rightarrow np + (n-1)p$).

\begin{figure}[htbp]
\begin{center}
\includegraphics[width = \columnwidth]{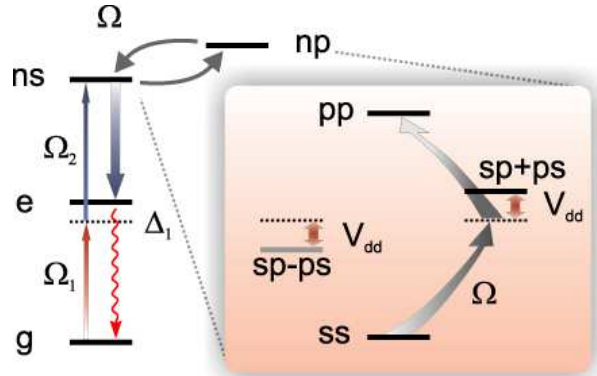}
\caption{Ground level atoms are two-photon excited to a Rydberg $s$-level, which is then mixed with a $p$-orbital by applying a microwave pulse of Rabi frequency $\Omega$. The inset shows the effect of dipole-dipole interaction and microwave dressing of an atom pair. The atomic spin-wave is retrieved from the Rydberg state with a $\pi$-pulse resonant to a low-lying excited state.}
\label{fig:coupling}
\end{center}
\end{figure}

We consider a cloud of cold alkali atoms. Since the procedure we propose is fast compared to atomic motional timescales in a cold ensemble, we assume that the positions of the atoms are fixed and postpone discussion of motional effects until later.
The relevant atomic levels are sketched in \fig{coupling}: the ground level $\ket{g}$, the first excited level $\ket{e}$, the target Rydberg level $\ket{s} = \ket{ns}$ and the Rydberg level used for the dephasing protocol $\ket{p_j} = \ket{n'p_{j}},$ $j=1/2,3/2$ to be discussed below. We suppress the state magnetic quantum numbers $m$ assuming that optical pumping of the ground level produces laser excitation of only a selected Zeeman state of the target Rydberg level; the coupling of magnetic sublevels by Rydberg interactions will be included in the results presented below.
While we consider a target $s$-orbital Rydberg level, it is possible to apply the formalism to $d$ levels as well.

\noindent \emph{Rydberg level excitation}:
The target Rydberg level $|ns\rangle$ is laser excited through a two-photon resonant transition with large single photon detuning $\Delta_1$. The effective excitation Rabi frequency is $\Omega_{2ph} = \Omega_1\Omega_2/(2\Delta_1)$ where $\Omega_1$ and $\Omega_2$ are the Rabi frequencies of the lasers (see \fig{coupling}).

The ensemble, defined by the region illuminated by the waist of the excitation laser (typically 50-60$\mu$m), contains $N \gg 1$ atoms. We may assume the atoms are initially independent since we consider (i) short laser pulses, whose bandwidth determines a Rydberg blockade radius \cite{saffman_review} that is much smaller than the size of the ensemble, and (ii) the pulse duration creates on average a single excitation, $T \approx 1/(\sqrt{N} \Omega_{2ph})$.
Thus laser excitation produces the state $\ket{\Psi_0} =  \sum_{\alpha = 0}^{N} c_{\alpha} \ket{\alpha_{\vv{k}_0}}$, where $\ket{\alpha_{\vv{k}_0}}$ contains $\alpha$ excitations with wavevector equal to the sum of the excitation laser wavevectors,  $\vv{k}_0 = \vv{k}_1 + \vv{k}_2$, and amplitude $c_{\alpha} = 1/\sqrt{e \alpha!}$ associated with a Poissonian distribution of unit mean. 
 The maximum efficiency of the single spin-wave preparation is $|c_1|^2 = 1/e$.
In the limit of weak excitation, $\ket{\alpha_{\vv{k}_0}} \approx (\hat{S}^{\dagger}_{\vv{k}_0})^{\alpha} \ket{0} / \sqrt{\alpha!}$, where $\hat{S}_{\vv{k}} = (1/\sqrt{N}) \sum e^{-i\vv{k}\cdot\vv{r}_{\mu}} \ket{g}_{\mu}\bra{s}$ is the spin-wave annihilation operator and the spin-wave vacuum $\ket{0}$ has all the atoms in the ground state: $\ket{\Psi_0}$ corresponds to a coherent state of spin waves.

\noindent \emph{Microwave-dressed Rydberg levels}:
Atoms laser excited to the Rydberg state $\ket{s}$ are assumed to have negligibly weak Van der Waals interactions.
A microwave field couples the transition $\ket{s}$-$\ket{p_j}$, transferring population to the $\ket{p_j}$ level and thus inducing a long-range resonant dipole-dipole interaction \cite{adamsmuwave} (see \fig{coupling}).

The interaction Hamiltonian for the atomic system is given by $\hat{H}_I = \sum_{\mu} \hat{V}_{\mathcal{E}}^{\mu} +  \sum_{\mu < \nu} \hat{V}_{dd}^{\mu\nu}$; the first term contains the electric dipole interaction of each atom $\mu$ with the microwave field: $\hat{V}_{\mathcal{E}}^{\mu} = - \hat{\vv{d}}_{\mu}\cdot \mathcal{E}(t)(\boldsymbol{\varepsilon} e^{-i \omega t} + \boldsymbol{\varepsilon}^{*} e^{i \omega t} )/2$,
where $\hat{\vv{d}}_{\mu}$ is the electric dipole moment of atom $\mu$ and $[\boldsymbol{\varepsilon}, \omega, \mathcal{E}(t)]$ are the polarization vector, angular frequency and time-dependent amplitude of the electric field. The Rabi frequency for the coupling between $\ket{s}$ and $\ket{p_j}$ is defined as $\hbar\Omega(t) = \mathcal{E}(t) \mathcal{D}$, where $\mathcal{D}$ is the reduced matrix element for the transition. The spatial phase of the microwave field is suppressed because the size of the sample is much smaller than its wavelength.
The dipole-dipole interaction between atoms is given by
\begin{equation}
\hat{V}_{dd}^{\mu\nu} = \frac{1}{4\pi\epsilon_0} \frac{1}{\mathcal{R}^3} \left[ \hat{\vv{d}}_{\mu} \cdot \hat{\vv{d}}_{\nu} - 3 \frac{(\hat{\vv{d}}_{\mu} \cdot \boldsymbol{\mathcal{R}}) (\hat{\vv{d}}_{\nu} \cdot \boldsymbol{\mathcal{R}})}{\mathcal{R}^2}  \right]
\end{equation}
where $\boldsymbol{\mathcal{R}}$ is the interatomic separation.
The dipole-dipole matrix element for a given angular momentum channel may be written as $C_3/\mathcal{R}^3$  \cite{saffman08, saffman_review},
and may be calculated with a semiclassical approach \cite{kaulakys}. We ignore non-resonant Van der Waals processes ($C_6/\mathcal{R}^6$) and retain only the resonant couplings \cite{gallagher_book,saffman08}.

\noindent\emph{Dephasing protocol}:
Resonant dipole-dipole interactions cause phase shifts of polarized Rydberg atom pairs, triples, etc.
The accumulated phase decouples these excitations from the phase matched radiation mode during the retrieval process.
With a suitable protocol, we can take advantage of this dephasing to generate high quality single photons.
A single channel model of the interaction is sufficient to demonstrate the physics of the protocol.

We consider a Ramsey-like $2\pi$-pulse sequence: a single-atom $\pi/2$ microwave pulse that polarizes the Rydberg atoms is followed by an interval $\Delta T$, during which resonant interactions occur and is terminated by a restoring $3\pi/2$-pulse.
A many-body state containing a single Rydberg excitation undergoes a complete rotation.
A given atom pair prepared in the target Rydberg level $\ket{s}$ experiences the following transformation assuming the regime of \emph{strong dressing}, $\Omega \gg V_{dd}$: (a) the $\pi/2$-pulse is responsible for the evolution $\ket{s}\ket{s} \rightarrow (1/2) [\ket{s} \ket{s} - \ket{p_j}\ket{p_j} + i (\ket{p_j}\ket{s} + \ket{s}\ket{p_j}) ]$, (b) during the interval $\Delta T$ the state transforms to $(1/2) [\ket{s}\ket{s} - \ket{p_j}\ket{p_j} + i\, e^{i \varphi} (\ket{p_j}\ket{s} + \ket{s}\ket{p_j})]$, where the phase $\varphi = V_{dd} \Delta T/\hbar$, (c) the $3\pi/2$-pulse $[\ket{s} \rightarrow 1/\sqrt{2} (- \ket{s} +i \ket{p_j}) $ and $\ket{p_j} \rightarrow 1/\sqrt{2} (i \ket{s} - \ket{p_j})]$, completes the overall transformation that maps $\ket{s}\ket{s}$ into
\begin{equation}
 \ket{\chi(\varphi)} = e^{i \frac{\varphi}{2}} \left[\ket{s}\ket{s} \cos\left(\frac{\varphi}{2}\right) + i \ket{p_j} \ket{p_j} \sin\left(\frac{\varphi}{2}\right) \right].
\label{eq:psi_toy}
\end{equation}
In the limit $\varphi \rightarrow 0$, we recover the initially prepared atom pair excitation $\ket{s}\ket{s}$ and more generally the survival amplitude for this state is $e^{i \varphi/2} \cos( \varphi/2)$.
The accumulated phase will be different for each atom pair. As a consequence, the probability for emission of a photon pair in the phase matched mode is reduced through destructive interference of distinct atom pair contributions, as we argue below. A small number of atom pairs will experience Rydberg blockade $\Omega \ll V_{dd}$ and while their effect is not detrimental it is anyway negligible.

We may write the many-body state after the Ramsey $2\pi$-pulse as $\ket{\Psi} = c_0 \ket{0} +  c_1 \ket{1_{\vv{k}_0}} + c_2 \ket{\Psi^{(2)}} + O(c_3)$, by defining
\begin{eqnarray}
\ket{\Psi^{(2)}}  = \sqrt{\frac{1}{\mathcal{N}}} \sum_{\nu >\mu=1}^{N-1} e^{i \vv{k}_0\cdot (\vv{r}_{\mu} +\vv{r}_{\nu})} \ket{\chi(\varphi_{\mu\nu})}_{\mu\nu}.
\label{eq:interactingpsi}
\end{eqnarray}
Here $\varphi_{\mu\nu}$ is the phase induced on the atom pair $(\mu,\nu)$, $\mathcal{N} = N(N-1)/2$ is the number of distinct pairs and $\ket{1_{\vv{k}_0}} = S^{\dagger}_{\vv{k}_0} \ket{0}$ is the one spin wave state. 

\noindent \emph{Optical readout procedure}:
Rydberg excitations in the $\ket{s}$ level can be optically retrieved by a laser of wavevector $\vv{k}_3$, transferring the population to an intermediate state which decays to the ground level as shown in \fig{coupling}.
The quality of single-photon emission is evaluated through measurement of the normalized correlation function, $g^{(2)} = \langle\hat{a}_{\vv{k}'_0}^{\dagger}\hat{a}_{\vv{k}'_0}^{\dagger}\hat{a}_{\vv{k}'_0}\hat{a}_{\vv{k}'_0} \rangle/\langle\hat{a}_{\vv{k}'_0}^{\dagger}\hat{a}_{\vv{k}'_0} \rangle^2$, in the phase matched direction, $\vv{k}'_0 = \vv{k}_0 - \vv{k}_3$. Here $\hat{a}_{\vv{k}'_0}$ is the annihilation operator for photons in the phase matched mode.
Analysis of the decay process shows that we may write $\hat{a}_{\vv{k}'_0} = \sqrt{\eta} \hat{S}_{\vv{k}_0} + \sqrt{1 - \eta} \hat{\xi}$, where the operator $\hat{\xi}$ describes vacuum fluctuations, and $\eta$ is the coupling efficiency.
Hence the correlation function becomes $g^{(2)} = \langle \hat{S}_{\vv{k}_0}^{\dagger}\hat{S}_{\vv{k}_0}^{\dagger}\hat{S}_{\vv{k}_0}\hat{S}_{\vv{k}_0} \rangle/ \langle\hat{S}_{\vv{k}_0}^{\dagger}\hat{S}_{\vv{k}_0}\rangle^2$.
The phase-matched retrieval of Poissonian excitations corresponds to the value $g^{(2)} =1$.
For simplicity, we focus on the effects of single and double excitations and truncate the many particle coherent state beyond this point: $c_{\alpha} = 0$ for $\alpha > 2$. The corresponding value of $g^{(2)}$ for the truncated coherent state is $g^{(2)}(0) = e/4$.
The state $\ket{\Psi}$ yields
\begin{equation}
g^{(2)} = \frac{4g^{(2)}(0) \abs{\frac{1}{N^2} \sum_{\mu =1}^{N} \sum_{\nu \neq \mu}  e^{i\frac{\varphi_{\mu\nu}}{2}} \cos\left(\frac{\varphi_{\mu\nu}}{2}\right)}^2}{ \left[ 1  +   \frac{1}{N^3} \sum_{\mu=1}^{N} \abs{\sum_{\nu \neq \mu} e^{i\frac{\varphi_{\mu\nu}}{2}}\cos\left(\frac{\varphi_{\mu\nu}}{2}\right)}^2\right]^2},
\label{eq:g2dephasing}
\end{equation}
in the spin-wave approximation $(N-1)/N \approx 1$.

This expression clearly shows that the probability for photon pair emission is a superposition of the atom pair contributions determined by the Ramsey protocol in \eq{psi_toy}: the broader the distribution of phases $\varphi_{\mu\nu}$ the more effective is the destructive interference of two-photon amplitudes.

Since the phases $\varphi_{\mu\nu}$ depend on time, we rewrite the correlation function of Eq.~\eq{g2dephasing} as $g^{(2)}(t) = 4g^{(2)}(0) f(t)/(1 + h(t))^2$, which defines $f(t)$ and $h(t)$. At $t=0$, it is trivial to verify that $f(t) = h(t) = 1$. For $t \rightarrow \infty$, assuming the atoms are randomly distributed in the ensemble, the accumulated phase behaves like a random variable: $\langle e^{i\varphi/2} \cos(\varphi/2)\rangle = 1/2$, which gives $f,h \rightarrow 1/4$. We therefore predict the asymptotic value of the correlation function after the Ramsey protocol to be $g^{(2)} \rightarrow g^{(2)}(0) 16/25$.

\begin{figure}[htbp]
\begin{center}
\includegraphics[width = \columnwidth]{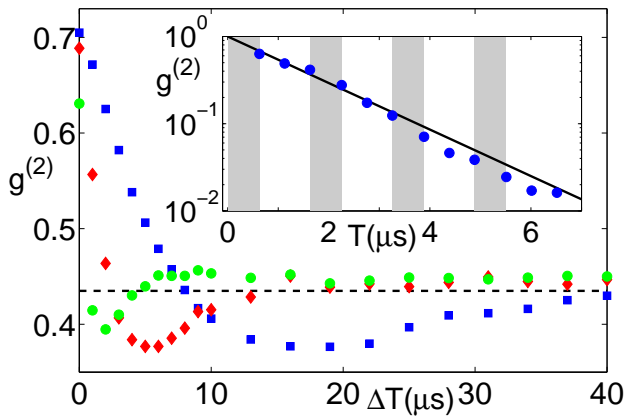}
\caption{Decay of the correlation function $g^{(2)}$ in the phase matched mode versus interaction time $\Delta T$ for a single Ramsey $2\pi$-pulse cycle in the strong dressing limit. Principal quantum numbers: $n = n' = 60$ (Blue squares), $n = n' = 79$ (Red diamonds) and $n = n' = 100$ (Green circles). Black dashed line represents the asymptotic limit $g^{(2)}(0)\;16/25$. Inset: effect of repeated cycles for $n = 100$ with $\Delta T = 1 \mu$s and $\Omega = 10^{7} \textrm{s}^{-1}$; first cycle $n' = 100$, $j= 1/2$, second cycle $n' =  99$, $j= 1/2$, third cycle $n' = 100$, $j= 3/2$ , fourth cycle $n' = 99$, $j= 3/2$. Dark regions correspond to the duration of each $2\pi$ microwave pulse cycle.}
\label{fig:g2}
\end{center}
\end{figure}

This asymptotic value agrees with simulations of a multichannel theory which also shows a stronger subpoissonian dip at short times, \fig{g2}.
As an example we take Rb comparing the time dependence for the Rydberg levels with principal quantum numbers $n = (60, 79,100)$.
We consider $N = 100$ atoms randomly distributed in a cubical box with side $L = 60 \mu$m.
The correlation function in Eq~\eq{g2dephasing} is found by solving the two-body Schrodinger equation for each atom pair.
The fine-structure transitions $\ket{ns}$-$\ket{n'p_{1/2,3/2}}$ have ($16$,$36$) available coupled channels involving different combinations of the magnetic quantum numbers, respectively.
While the asymptotic value of $g^{(2)}$ is independent of $n$, we observe that the rate of dephasing, as measured by the temporal position of the minimum, scales in inverse proportion to the strength of the interaction $V_{dd} \propto n^4$.

\noindent\emph{Single photon generation}:
Although a single Ramsey $2\pi$-pulse cycle produces subpoissonian emission statistics asymptotically,
the more pronounced and fast initial dephasing of $g^{(2)}$ suggests an improved protocol in which i) the value of $\Delta T$ optimizes the transient dephasing, and ii) $R$ repetitions of the Ramsey cycle further reduce $g^{(2)}$ by the replacement
\begin{equation}
e^{i\frac{\varphi_{\mu\nu}}{2}} \cos\left(\frac{1}{2}\varphi_{\mu\nu}\right) \rightarrow \prod_{q = 1}^{R} e^{i\frac{\varphi^{(q)}_{\mu\nu}}{2}} \cos\left(\frac{1}{2}\varphi_{\mu\nu}\right)
\label{eq:acc_phase}
\end{equation}
in Eq.~\eq{g2dephasing}.
In each of the repeated cycles, it is essential that the microwave field couples the $\ket{ns}$ target level to a different and unpopulated $\ket{n'p_j}$ level otherwise the coherence established with the target level invalidates Eq.~\eq{acc_phase}.
As each cycle represents a single-particle $2\pi$-pulse the single photon contribution is unaffected throughout.

For $n \gg 1$, the channels $\ket{ns} \leftrightarrow \ket{np_j}$ and $\ket{ns} \leftrightarrow \ket{n-1,p_j}$ have similar interaction strengths and may each be employed in the repeated Ramsey protocol.
In \fig{g2} inset, we show a sample evolution of $g^{(2)}$ for a sequence of $R = 4$ Ramsey cycles with $n = 100$ and $j = 1/2, 3/2$. We plot the function $e^{-T/\tau}$ where $\tau = \Delta T + 2\pi/\Omega$ is the duration of a single cycle.
The comparison shows that the decay rate of $g^{(2)}$ is approximately exponential and high quality single excitations are generated in several $\mu$s, well within the lifetime of this Rydberg state.
For comparison, the DLCZ protocol \cite{DLCZNAT2001,pan_2007,*vuletic2007,*Pan2008}, which has a typical trial period $\sim 1  \mu$s, and excitation probability $\sim 10^{-3}$, generates single excitations in a time of order $1$ ms \cite{RZhao_2009}.

\noindent \emph{Entanglement of Rydberg spin waves}:
We have shown that the repeated Ramsey protocol allows the fast creation of a single spin wave, stored in the Rydberg level $\ket{ns}$, coupled to the phase matched mode. An independent spin wave associated with the orbital $\ket{n's}$ can also be generated provided the dipole-dipole interaction does not cause interference of the dephasing protocols for the levels $n$ and $n'$. For $n \gg 1$, the dipole coupling between Rydberg levels $ns$ and $n'p$ decays rapidly as the energy difference increases: for example, for $n = 100$ the interaction strength between adjacent $100s$ and $99p_j$ or $100p_j$ orbitals is $100$ times larger than that for $100s$ and $98p_j$ or $101p_j$ \cite{kaulakys,saffman_review}.  These single spin waves may also be transported through the Rydberg spectrum by using single-particle microwave $\pi$-pulses: the phase matching condition is not affected provided the Rydberg transition wavelength is much larger than the ensemble size.

\begin{figure}[htbp]
\begin{center}
\includegraphics[width = \columnwidth]{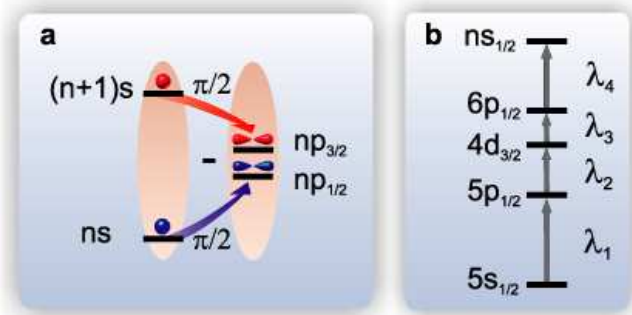}
\caption{a) Atomic levels involved in entanglement of two spin-wave excitations via a protocol described in the text.
b) Four-photon excitation of Rydberg spin-waves in atomic Rb, with $(\lambda_1, \lambda_2,\lambda_3,\lambda_4)=(795,1475, 2294, 1005)$ nm.  Collinear and off-axis geometries lead to spin waves of period $50$ $\mu$m, and $\infty$, respectively.}
\label{fig:gate}
\end{center}
\end{figure}

A protocol to entangle two independent spin waves in levels $n$ and $n' = n+1$ is sketched in \fig{gate}.
We define the operators $\hat{S}^{\dagger}_{n}$ and $\hat{P}^{\dagger}_{nj}$ which create spin waves in the $\ket{ns}$ and $\ket{np_j}$ orbitals, respectively.
A pair of strong $\pi/2$-pulses couple the $\ket{ns}$ and $\ket{(n+1)s}$ levels to $\ket{np_{1/2}}$ and $\ket{np_{3/2}}$ respectively, transforming $\hat{S}^{\dagger}_{n+1}\hat{S}^{\dagger}_{n} \ket{0} \rightarrow [\hat{S}^{\dagger}_{n+1}\hat{S}^{\dagger}_{n} \ket{0} - \hat{P}^{\dagger}_{n3/2}\hat{P}^{\dagger}_{n1/2} \ket{0} + i (\hat{S}^{\dagger}_{n+1}\hat{P}^{\dagger}_{n1/2} \ket{0} + \hat{P}^{\dagger}_{n3/2}\hat{S}^{\dagger}_{n} \ket{0}) ]/2$.
The resonant dipole-dipole interactions couple the $s$ and $p$ orbitals thus inducing different phase shifts for each atom pair.
The system evolves into the many body state $\ket{0} + c_{1,0}/2 (\hat{S}^{\dagger}_{n+1} + \hat{S}^{\dagger}_{n} + i \hat{P}^{\dagger}_{n1/2} +i \hat{P}^{\dagger}_{n3/2} ) \ket{0} + c^2_{1,0} \ket{\Phi} $ with
\begin{equation}
\ket{\Phi} = \frac{1}{2}\left[ \hat{S}^{\dagger}_{n+1}\hat{S}^{\dagger}_{n} \ket{0} - \hat{P}^{\dagger}_{n1/2}\hat{P}^{\dagger}_{n3/2} \ket{0} + \frac{i}{N} \sum_{\mu,\nu} \ket{\xi_{\mu\nu}}\right],
\end{equation}
$\ket{\xi_{\mu\nu}}  = e^{i \vv{k}_0 \cdot (\vv{r}_{\mu} + \vv{r}_{\nu})} [e^{i \varphi'_{\mu\nu}}\ket{(n+1)s}_{\mu}\ket{np_{1/2}}_{\nu} + e^{i \varphi_{\mu\nu}} \ket{ns}_{\mu}\ket{np_{3/2}}_{\nu}]$ and $c_{1,0} = c_1/c_0$.
As for the single photon generation, the phase matched retrieval process causes the suppression of the contribution of the states $\ket{\xi_{\mu\nu}}$ as result of the destructive interference of the atom pair amplitudes.
The state $\ket{\Phi}$ may be mapped onto long-lived ground state coherences \cite{saffmannature,*grangiernature} or optically retrieved, via $s$ or $d$ states, resulting in entangled polarization or time-bin qubits.
Entangled pairs of long-lived qubits, or the atom-light qubit pair, can be used as building blocks in larger entangled systems, for example, in entanglement purification protocols \cite{DLCZNAT2001,CollinsEtAlPRL2007,*Jiang2007}.

The motional dephasing of optical ground-Rydberg coherences has been a serious problem in exploiting Rydberg atom interactions \cite{saffmannature,grangiernature}. While atom trapping would in principle alleviate this effect, so far no effective Rydberg atom confinement schemes have been demonstrated, although there are promising works in that direction \cite{raithel2010, *Zhang2011}. For a cold MOT of Rb the average atomic velocity $v\sim 0.1$ m/s, while the spin wave grating period for two-photon excitation is only $\Lambda \sim 1$ $\mu$m, giving a coherence time $\Lambda/(2\pi v) \sim 2$ $\mu$s.
In order to overcome this limitation, we propose the four-photon excitation scheme shown in \fig{gate}(b) for atomic Rb. In this case the four wavevector mismatch can be made equal to zero and the corresponding spin wave period diverges, thereby eliminating motional decoherence.
Since all the transitions involved possess strong dipole moments, the Rabi frequency can easily exceed several MHz with available laser powers.
An optical depth of $10$ is achievable with a gas of density $10^{12}$ cm$^{-3}$ and diameter $50$ $\mu$m, sufficient for photon retrieval in the phase matched mode \cite{duan2002}.
The maximum efficiency of single spin-wave preparation, and hence single-photon generation, is given
by $1/e$. 

In conclusion, we propose techniques for the fast creation of single quantum excitations and entanglement of such excitations in large atomic ensembles, suitable for efficient light-matter state transfer.
The protocols we propose are based purely on the dephasing of multiple excitations due to resonant $1/r^3$ dipole-dipole interactions induced by microwave coupling of opposite parity Rydberg states. 
In the future, it will be interesting to investigate an intermediate regime between blockade and dephasing that optimizes efficiency, speed and error probability for laboratory implementation.

\noindent We thank AFOSR and NSF for support.
\bibliography{references}

\end{document}